# The elastic constants of $MgSiO_3$ perovskite at pressures and temperatures of the Earth's mantle.

Artem R. Oganov, John P. Brodholt, G. David Price

*Department of Geological Sciences, University College London, Gower Street, London WC1E 6BT, U.K.*
Tel.: +44-(020)-7679-3344; fax: +44-(020)-7387-1612.

Convection in the Earth's mantle is fundamentally responsible for plate tectonics, and the associated continental drift, earthquakes, and volcanism[1]. The convective temperature anomalies, associated with the ascending hot and descending cold streams, can be obtained from seismic tomography, provided elastic properties of the mantle minerals are known as a function of temperature at mantle pressures (up to 1.36 Mbar). Such information cannot currently be given by experiments. Circumventing this experimentally formidable task, we perform *ab initio* molecular dynamics (*AIMD*) simulations[2,3] of elastic and seismological properties of $MgSiO_3$ perovskite, the major mineral of the lower mantle (LM), at relevant thermodynamic conditions. We believe that these are the first true finite-temperature *AIMD* simulations of elastic constants to be performed for any material. Our results imply that the LM is either significantly anelastic[4] or compositionally heterogeneous on the large scale[5]. The temperature contrast between the cold slabs and hot plumes was found to be about 800 K at 1,000 km, increasing to 1,500 K at 2,000 km and, possibly, over 2,000 K at the core-mantle boundary.

*Ab initio* molecular dynamics (*AIMD*), pioneered by Car and Parrinello[2], has made an enormous impact on modern planetary and materials sciences[6-8]. Advantages of this approach include the explicit account of the quantum nature of electronic structure and interatomic interactions, and the description of intrinsic anharmonicity of lattice vibrations. Although extremely computationally expensive, this is the method of choice for simulating high-temperature significantly anharmonic phenomena, such as melting[8,9], ionic conductivity, displacive phase transitions, thermal expansion[10], and elastic properties. In the past the outstanding computational expense precluded the use of *AIMD* for studying the elasticity of materials. Calculations of elastic constants using this method can be done conveniently via one of two routes: 1) from fluctuations of stress or strain, 2) from stress-strain relations, i.e. from Hooke's law. Both routes require enormous computational efforts because very long runs are required by the first method, or in the case of the second method because of the necessity to perform several independent *AIMD* simulations for different lattice strains that include the non-linearity of stress-strain relations. Recent advances in computer technology and increased accessibility of supercomputers have for the first time now made such simulations possible.

Orthorhombic $MgSiO_3$ perovskite is an ideal subject for *AIMD* simulations. When containing some Fe, it is the most abundant mineral in the Earth, comprising 60-100 vol.% of the LM and dominating many of its properties. The LM comprises over 50% of the Earth's volume; extending between the depths of 670 km and 2,891 km, it is characterised by very high pressures (24-136 GPa) and temperatures, roughly between 2,000 K and 3,000 K, possibly rising to *ca* 4,000 K[11]. At these temperatures atomic motion is essentially classical and possibly significantly anharmonic, making molecular dynamics approach well justified. A hypothetical cubic phase with superionic conductivity has been proposed as an explanation of the observed high electrical conductivity of the lower mantle[12,13]; molecular dynamics is the only currently viable approach to simulate fast ionic conduction phenomena in solids. Our previous work[14] considered static (i.e. neglecting thermal motion of atoms) elastic constants and employed *AIMD* to study thermal expansivity and thermal equation of state of $MgSiO_3$ perovskite, which was found to stay orthorhombic at LM temperatures. Here we perform *AIMD* simulations of elastic constants of $MgSiO_3$ perovskite at LM conditions with a full account of temperature. At this stage we consider only the pure $MgSiO_3$ perovskite. The effects of the moderate Fe content, while potentially important for certain properties (e.g., shear modulus), are expected to be negligible for others, especially for the thermal expansion, bulk modulus, and *P-T*-derivatives of the elastic moduli and seismic wave velocities, which are the main subject of this work.

Seismic tomography gives an invaluable insight into the 3D-structure and dynamics of the Earth, but its quantitative interpretation requires knowledge of temperature variations of seismic wave velocities. Seismology gives $\nu_T=(\frac{\partial \ln V_S}{\partial \ln V_P})_T$, the ratio of variations of the shear ($V_S$) and compressional ($V_P$) velocities due to pressure alone, to be 0.7[1]. A similarly defined parameter, $\nu_P=(\frac{\partial \ln V_S}{\partial \ln V_P})_P$, measuring the same ratio, but due to temperature effects alone, is much larger: it increases from 1.7 to 2.6 between the depths of 1,000 km and 2,000 km according to seismic tomography data[15]. The large difference between $\nu_T$ and $\nu_P$ was a puzzle for geophysicists over last 15 years; D.L. Anderson[1] suggested that the difference is entirely due to intrinsic anharmonicity, to simulate which one needs to go beyond the quasiharmonic approximation and use methods such as molecular dynamics or Monte Carlo.

Our *AIMD* simulations (see legend to Fig. 1 and Fig. 2) were performed with the VASP code[2] run on 128 nodes of the CRAY T3E supercomputer at Manchester Computer Centre. The calculated equilibrium lattice parameters, isothermal and adiabatic elastic constants, and thermal expansion coefficients are presented in Table 1. One of the calculated points, 38 GPa and 3500 K, is close to the melting of $MgSiO_3$ perovskite[21]; although we observed very large displacements of atoms, all atoms vibrated about their ideal crystallographic locations and did not diffuse, and shear elastic constants were large and positive. It is remarkable that even at these conditions orthorhombic $MgSiO_3$ perovskite does not transform to a cubic or tetragonal phase. Anomalous elastic constants obtained at these critical conditions were not used in the further analysis.

We calculate $\nu_T$=0.61-0.81 (our static simulations[14] give 0.59-0.74) in keeping with geophysical observations. The $\nu_P$ increases from 1.5 at the depth of 1,000 km to 1.9 at 2,000 km. These values, obtained with full account of anharmonic effects, are still somewhat lower than the observed values. Karki *et al.*[22], using the quasiharmonic approximation, found $v_P$ of MgO to increase from 1.4 to 1.9 from to top to the bottom of the LM. This agrees well with our results for $MgSiO_3$; the still remaining deficit of $v_P$ can only be explained by either significant anelasticity[4] or large-scale compositional heterogeneity of the LM[5].

Seismic tomography is a quickly developing technique; the most recent tomography maps have similar qualitative features (locations of cold slabs and hot plumes) and give the absolute velocity perturbations within ~25% uncertainty. For interpreting seismic tomography in terms of temperature it is most convenient to use the bulk sound velocity $V_\Phi$, whose seismological determination is unaffected by anelastic effects. The bulk velocity maps can be accurately determined by the joint inversion of the shear and compressional velocities. Perhaps, the most reliable global seismic tomography maps currently available are those of Masters *et al.*[5]. Older higher-resolution maps of Kennett *et al.*[23] realistically give much narrower cold (high-velocity) zones, but strongly underestimate amplitudes of the velocity variations.

We find the temperature derivative of the bulk velocity, $\phi=(\partial \ln V_\Phi/\partial T)_P$, to vary from -$1.7*10^{-5}$ $K^{-1}$ at 1,000 km to $-0.9*10^{-5}$ $K^{-1}$ (errors $\pm 20\%$) at 2,000 km depth. Temperature variations can be obtained from the relative velocity perturbations: $\delta T=(\delta V_\phi/V_\phi)/\phi$. Combined with bulk sound velocity maps of Masters *et al.*[5] (maximum velocity contrast $\frac{\Delta V_\phi}{V_\phi}= 1.4\%$ at 1,225 km and 2,195 km), this gives the temperature contrast between the hot plumes and cold slabs increasing from 800 K to 1,500 K between 1,000 and 2,000 km. Linear extrapolation gives ~550 K at the top and 2,150 K at the bottom of the LM. The root-mean square temperature variations are ~150-250 K across the LM. Our temperature contrasts are ~2-4 times smaller than estimates[24] obtained using extremely uncertain extrapolations of $s=(\partial \ln V_S/\partial T)_P$ and which would suggest partial melting of the LM. Our results can support partial melting only in the lowermost part of the LM. Castle *et al.*[25], using their shear velocity maps for the bottom of the LM (maximum $\frac{\Delta V_s}{V_s}$ = 7.4%) and a reasonable guess for *s*, obtained an estimate of the temperature contrast at the core-mantle boundary similar to ours, ~1,600 K. Using the $V_\phi$-maps of Kennett *et al.*[23] (maximum $\frac{\Delta V_\phi}{V_\phi}$ = 1.0% at 1,225 km and 2,195 km), we get much smaller temperature contrasts, rising from ~500 K to ~1,000 K between the depths of 1,000 km and 2,000 km.

It is becoming increasingly clear that temperature variations are the main factor determining the perturbations of seismic wave velocities in the LM. Anelasticity can be important for the shear velocities, but is negligible for the bulk velocities. Compositional heterogeneity is significant only near the core-mantle boundary[5]; it can occur due to the chemical reaction[26] between the core and mantle, driving Fe from the silicate mantle into the metallic core. Hot plumes, rising from the core-mantle boundary, would then be depleted in Fe. When substituting for Mg, Fe dramatically decreases shear moduli of many silicates and oxides, while the effect on bulk moduli is very small. Hence, even minor regional variations of the Fe content can have a major effect on the shear velocities, being less important for the bulk velocity. E.g., very hot, slightly Fe-depleted and very cold, slightly Fe-enriched mantle rocks can have the same shear velocities. Consistent the depletion of plumes in Fe, we find the temperature contrasts at 2,000 km determined from the shear velocities[5,23] (1,000 K from maps[5]) to be smaller than those determined from the corresponding bulk velocities (1,500 K from maps[5]). This effect can even produce an anticorrelation between the bulk and shear velocities, but not its observed[5] pattern: it remains a mystery why it is the bulk (not shear) velocity anomalies that undergo a reversal near the core-mantle boundary (i.e. low-velocity zones at the core-mantle boundary underlay high-velocity anomalies of the rest of the LM). In any case, the temperature contrasts obtained from the bulk velocities must be the most reliable, at least outside the anomalous core-mantle boundary region, where compositional effects seem to be significant for all types of seismic waves. The temperature contrasts that we have found can be used as important constraints in numerical modelling of the mantle convection.

Kesson *et al.*[27] estimated that lithospheric slabs should be at least ~650 K colder than surrounding mantle if they are to sink to the core-mantle boundary and at least 250 K colder to reach the depth of 1,100 km. Our maximum cold temperature anomalies (roughly half of the total temperature contrasts) are similar to these estimates and suggest that some lithospheric slabs might stop sinking before reaching the core-mantle boundary. Neutrally buoyant slabs will be dissolved by the convecting mantle; some tomographic maps show most slabs disappearing in the middle of the LM[23], while others[5] show that most slabs do reach the core-mantle boundary.

The next important step is to estimate the extent of chemical heterogeneity and anelasticity in the LM and construct a 3D- mineralogical model of the Earth's mantle. With improved seismic tomography models and mineral physics data this can be achieved in foreseeable future. We believe that *AIMD* simulations will play an important role in solving this and many other geologically important problems.

**Acknowledgements**

We gratefully acknowledge the receipt of the Russian President Scholarship for Education Abroad, UCL Graduate School Research Scholarship, and UK Overseas Research Scholarship (ARO) and of a Royal Society University Research fellowship (JPB). We acknowledge NERC for the access to the supercomputer facilities.

---

Correspondence and requests for materials should be addressed to A.R. Oganov (a.oganov@ucl.ac.uk)

Table 1. Properties of MgSiO$_3$ perovskite at high pressures and temperatures[a,b]

| Property | 38 GPa, 1500 K | 38 GPa, 2500 K | 38 GPa, 3500 K | 88 GPa, 1500 K | 88 GPa, 3500 K |
|---|---|---|---|---|---|
| $V$ | 147.63 | 150.83 | 154.08 | 132.55 | 136.48 |
| $a$ | 4.628 | 4.675 | 4.723 | 4.435 | 4.499 |
| $b$ | 4.792 | 4.814 | 4.836 | 4.655 | 4.674 |
| $c$ | 6.658 | 6.703 | 6.747 | 6.420 | 6.492 |
| $\sigma_{1,2,3}$ | 37.9; 38.0; 37.9 | 38.0; 37.6; 37.3 | 37.0; 39.0; 38.0 | 88.1; 87.7; 88.4 | 87.8; 87.9; 88.5 |
| $P$ | 37.9 | 37.6 | 38.0 | 88.1 | 88.1 |
| $C_{11}^{T/S}$ | 601 /616 | 553 /579 | 525 /564 | 813 /826 | 749 /783 |
| $C_{22}^{T/S}$ | 697 /711 | 629 /654 | 545 /578 | 978 /991 | 847 /878 |
| $C_{33}^{T/S}$ | 649 /663 | 591 /615 | 545 /580 | 933 /945 | 821 /850 |
| $C_{12}^{T/S}$ | 266 / 280 | 233 /258 | 222 /258 | 464 /477 | 398 /431 |
| $C_{13}^{T/S}$ | 235 /249 | 218 /243 | 218 /256 | 348 /362 | 324 /356 |
| $C_{23}^{T/S}$ | 251 /264 | 240 /265 | 251 /286 | 383 /396 | 364 /395 |
| $C_{44}$ | 262 | 232 | 202 | 336 | 270 |
| $C_{55}$ | 219 | 210 | 180 | 266 | 234 |
| $C_{66}$ | 199 | 178 | 147 | 264 | 195 |
| $K^{T/S}$ | 382.7 /396.8 | 349.9 /375.8 | 332.7 /369.5 | 565.7 /579.4 | 508.6 /542.5 |
| $G$ | 214.3 | 194.9 | 166.2 | 269.7 | 226.9 |
| $\gamma$ | 1.30 | 1.37 | 1.40 | 1.21 | 1.26 |
| $\alpha$ | 1.91 | 2.15 | 2.26 | 1.34 | 1.51 |

[a] - $V$, $a$, $b$, $c$ are unit cell volume and lattice parameters, in Å$^3$ and Å, respectively; $\sigma_{1,2,3}$, are stress tensor components, showing that non-hydrostatic deviations from the average pressure $P$ are very small; $C_{ij}$ are elastic constants (*T*- isothermal and *S*- adiabatic) in GPa (errors of a few %); *K* and *G* are Voigt-Reuss-Hill bulk and shear moduli in GPa (errors within 2%); $\gamma$ is the Grüneisen parameter and $\alpha$ thermal expansion in 10$^{-5}$ K$^{-1}$ (errors within 10%).

[b] - Important derivatives: $(\partial \ln V_S/\partial T)_P$=-3.67*10$^{-5}$ K$^{-1}$ (38 GPa), -3.78*10$^{-5}$ K$^{-1}$ (88 GPa). $(\partial \ln V_P/\partial T)_P$=-2.49*10$^{-5}$ K$^{-1}$ (38 GPa), -1.98*10$^{-5}$ K$^{-1}$ (88 GPa). $(\partial \ln V_\Phi/\partial T)_P$=-1.65*10$^{-5}$ K$^{-1}$ (38 GPa), -0.92*10$^{-5}$ K$^{-1}$ (88 GPa). $(\partial \ln V_S/\partial \ln V_P)_T$=0.61 (1500 K), 0.81 (3500 K). $(\partial \ln V_S/\partial \ln V_P)_P$=1.48 (38 GPa), 1.91 (88 GPa). All errors are within 20%.

**FIGURE LEGEND:**

FIG. 1. Snapshot of the crystal structure of MgSiO$_3$ perovskite from *ab initio* molecular dynamics simulations at 88 GPa and 3500 K. The equilibrium structure at these conditions is orthorhombic. *Spheres* are Mg atoms, *polyhedra* – SiO$_6$ octahedra. The supercell used in calculations (2*a*x2*b*x1*c* *Pbnm*-supercell, contours shown) contains 80 atoms. Simulations were also performed at 38 GPa and 1500 K, 38 GPa and 2500 K, 38 GPa and 3500 K, 88 GPa and 1500 K, and 136 GPa and 4500 K, and at all conditions the optimised cell was orthorhombic. The simulated pressures of 38 GPa and 88 GPa correspond to the depths of approximately 1,000 km and 2,000 km, respectively. Our simulations are based on density functional theory[16,17] within the generalised gradient approximation[18] and pseudopotential plane wave scheme. The pseudopotentials used are non-local, ultrasoft for O and norm-conserving with partial core corrections for Mg and Si. Plane-wave cutoff of 500 eV, which gives excellent convergence of all properties with respect to the basis set size, was used. Supercell periodicity was imposed on the wavefunction, which proved to give sufficiently accurate results with our large supercell. At each conditions the simulations were run in the constant *NVT* ensemble[19] for at least 0.79 ps after equilibration with 1 fs timestep. This was sufficient to produce accurate statistical averages of the stress tensor components. Cell dimensions were carefully optimised by making the stress tensor hydrostatic.

FIG. 2. Fluctuations of the stress tensor components for the structure shown in Fig. 1. Data for the optimised cell at 88 GPa and 3500 K show that the stress is hydrostatic. As the generalised gradient approximation often results in a shifted pressure scale[14], we corrected the calculated pressures and diagonal stress components by –12.08 GPa. This correction alters only the pressure scale, but not the properties. More information on the pressure correction and performance of the generalised gradient approximation can be found in[14]. Applying strains (we used one triclinic and three axial strains with magnitudes

+0.02 and –0.02) and calculating the time-averages of the induced stresses, we calculated the isothermal elastic constants $C_{ij}$ from non-linear stress-strain relations; different elastic constants are obtained from different stresses. Adiabatic elastic constants, used in calculations of seismic velocities, were calculated from the isothermal constants[20] and the thermal pressure tensor calculated directly by *AIMD*.

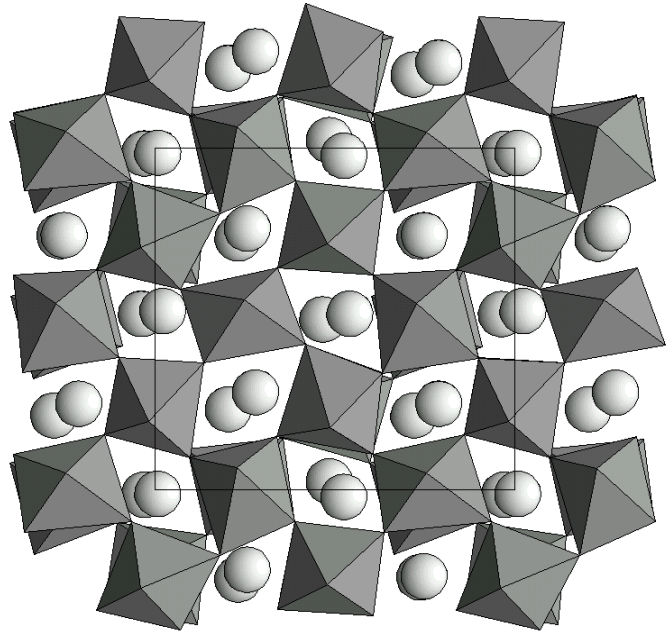

FIG. 1

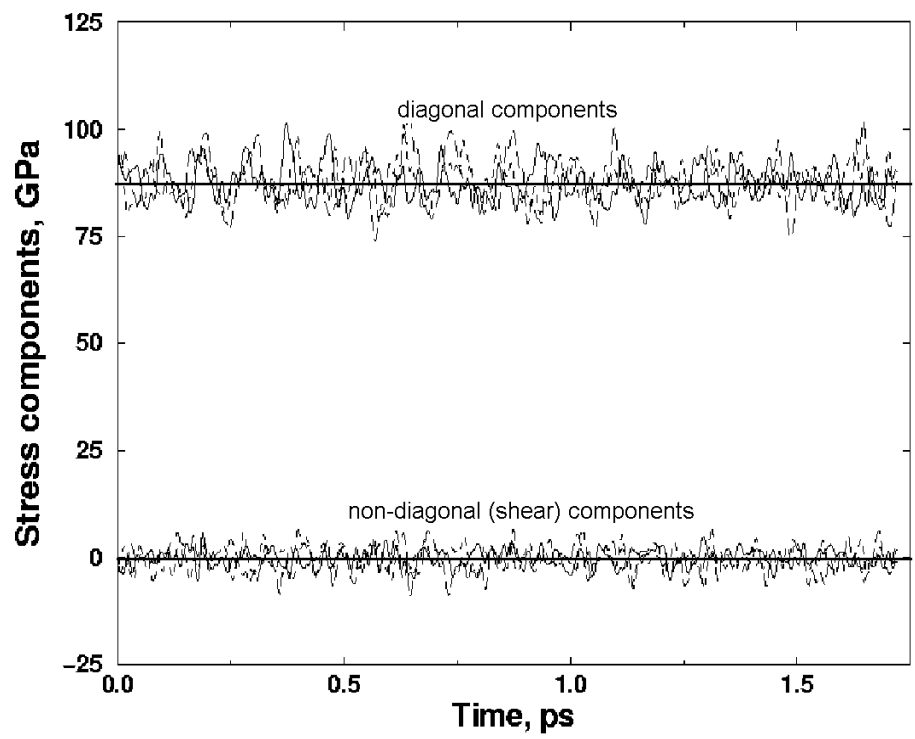

FIG. 2